%Paper: hep-th/9305159
%From: Boris Spokoiny <BORIS%JPNYITP.BITNET@pucc.Princeton.EDU>
%Date: Fri, 28 May 93 16:56:16 JST
%Date (revised): Fri, 04 Jun 93 21:55:58 JST
%Date (revised): Sat, 05 Jun 93 16:27:54 JST

\input phyzzx
\endpage
\centerline{ Stochastic  non de Sitter Inflation \dag}\footnote{\dag}{
 published in the Proceedings of the International Symposium on
Quantum Physics and the Universe, Waseda Univ. Tokyo Aug.19 - 22 (1992)
(Vistas in Astronomy,vol.37,pp.481-485,1993)}

\centerline{Boris~ Spokoiny \ddag}\footnote{\ddag}{E-mail BORIS@YITP.bitnet
 \endline  Address after June 23,~1993: ~Landau Institute for Theoretical
Physics, Russian Academy of Sciences, 142432 Chernogolovka, Moscow region,
Russia.}

The stochastic approach to quasi-de Sitter Universes (Starobinsky
(1986), Sasaki et al.(1988), etc.) appeared to be very fruitful and so it seems
In the chaotic inflationary model it is supposed that the evolution of the
Unive
condition$ (logV(\phi))'\ll M^{-1}_{pl}, $
which provides the slow rolling down of the scalar field along the
 potential and quasi-exponential evolution of the Universe.In this note
 we generalize the stochastic approach to the models which do not
 satisfy this restriction .We weaken this restriction and suppose that
$$
 (logV(\phi))''\ll M^{-2}_{pl}. \eqno(1)
 $$
So our models include quasi-exponential potentials
$  logV(\phi)=\lambda\phi/M +log{\tilde V}(\phi)  $
with a slowly varying $  log{\tilde V}(\phi) $. Such potentials give rise to
the
  coefficient) which was not calculated by them.So we think that the detailed
in

First we consider the theory of scalar perturbations in the model with a pure
ex
     $$
 L=-{1\over2}M^2R +{1\over2}(\partial_{\mu}\phi)^2-
{1\over2}V_0\exp{(\lambda\ph
 \eqno(2)
  $$
\v
The equations for the homogeneous background have a power-law solution (Lucchin
\vskip-7mm
 $$
a(t)=t^{n+1},~~~~ \phi=\varphi(t)=\varphi_0 - \sqrt{2(n+1)}M\log t, \eqno(3)
$$
\vskip-3mm
\noindent
where $ n={2/{\lambda^2}}-1 $. Solution (3)is an attractor solution of
the system (2) (see Halliwell (1987), Yokoyama \& Maeda (1988)).
In the longitudinal gauge we may represent the action for the scalar
perturbations (of the scalar field and metric) in the form
\vskip-5mm
$$
 S=2M^4k^2\int d{\eta}[{\chi'_k}^2- {(k^2 - 2({a'/ a})^2 + {a''/ a}
-2{({\varphi
   \eqno{(4)} ,
 $$
\vskip-3mm
\noindent
where k is a wave vector,$ \eta $ is a conformal time,
 $ {\chi}_k ={ h_k / {\dot\varphi}} $ , $h_k $ is a perturbation of metric.
Eq
                                                            $$ \chi_k = -
{\sqrt
$$    \vskip-3mm
\delta\phi_k=-2M^2({\varphi''/ \varphi'a})\chi_k -2M^2({\chi'_k/ a})    $
 \delta\phi_k = -{i\over {\sqrt2}}{H_*^{1+{1\over n}}\over k^{{3\over 2} +
{1\ov
  $$
wh
\vskip-5mm
$$
 H(\eta) = (1+{1\over n})H_*(-H_*\eta)^{1\over n},
 $$
$
      F(n)= {2^{1\over n}\sqrt{\pi}(1 + {1\over n})\over cos{\pi\over
n}\Gamma({
 $$
 For the de Sitter space $ F=F(\infty)=1$ The perturbation of the metric at $
-k
 $$
 h_k = -{iF(n)\over 2\sqrt{n+1}}{H_*^{1+{1\over n}}\over Mk^{{3\over 2}+{1\over
   $$
Now we want to work in the synchronous coordinate system:
$
  $$
 \delta\phi_s(k,t) = \delta\phi_k + {\dot\varphi}T_k,~~~~~
     \eqno(9)
 $$
\vskip-5mm
The above transformation of time is correct for long wave perturbations with  $
as a sum of a long wave part $ f(t,\vec r) $   (with $ k_f = {k\over a}<
\epsilo
 $$ F = f(t,\vec r) + \int {d^3k\over {(2\pi)^{3\over 2}}}\theta(k- \epsilon
aH)
where  the perturbation $ \hat f_s(k,t) $ under the integral is
 given in a synchronous coordinate system and satisfies the equations
 of motion for perturbations.We write down five equations similar
 to eq.(10) where we substitute triples
\nextline
 $ (\phi;\varphi(t,\vec r);{\hat \phi_s(k,t)}),$ ~
 $ (\dot{\phi};v(t,\vec r);\dot{\hat\phi}_s(k,t)),$ ~
$ (\ddot{\phi};w(t,\vec r);\ddot{\hat\phi}_s(k,t)), $
\nextline
$ (\alpha;\alpha(t,\vec r);\hat\alpha_s(k,t)),  $
$ (\dot\alpha;H(t,\vec r);\dot{\hat\alpha}_s(k,t))$
into eq.(10) instead of the triple
\nextline
$ (F;f(t,\vec r);\hat f_s(k,t))$. Here
\vskip-10mm
$$ \alpha(t,\vec r) = \log a(t,\vec r),     $$
 \vskip-6mm
The equation for the scalar field and the $ _o^o $ -Einstein equation after the
$ S_o + S_1 + S_2 = 0,  $
where $ S_0 $ is the contribution of the long wave part only,~$ S_1 $ is linear
\vskip-6mm

$$ w + 3Hv + V'= 0,~~~~~H^2 = {{({{v^2}/2} + V(\varphi))}/ 3M^2}.\eqno(11) $$
\vskip-5mm
After introducing a new variable $ \vartheta = {v/ H} $ we obtain
 from (10)-(11) after some calculations
$$
 {d\varphi\over d\alpha} = \vartheta + {1\over H}\epsilon(aH)\dot{}\int
{d^3k\ov
 $$
$$
 {d\vartheta\over d\alpha} = -(3-{\vartheta^2\over 2M^2})(\vartheta +
M^2{V'\ove
 \eqno(13)
 $$
where
$$
 \xi_k = \phi_s - {(\dot\varphi/ H)}\alpha_s = \delta\phi_k - {(\dot\varphi/
H)}
\eqno(14)
 $$
is a gauge invariant quantity and after the substitution of (3),(6) and (8)
into
important result since it shows that we have chosen good variables  ($
\vartheta
$
 \delta_1 \propto max(\epsilon^2,{1\over n} \epsilon^{1 + {2\over n}})
 $
due to $ \delta(k - \epsilon aH) $. So we may neglect this contribution to $
\ch
If we take into account the difference between a real quasi-power law
inflationa
$
 \delta_2 = O({(logV)''\over {M_{pl}}^2}) $
which corresponds to the factor $ m^2\over H^2 $ in the usual quasi-de Sitter
mo
So we put $ \chi = 0 $.

A general solution to the eq.(13) (with $ \chi = 0 $) contains two modes.
As usual we consider only the slowly varying mode
\vskip-5mm
$$
 \vartheta = -M^2(logV)'
 \eqno(15)
 $$
\vskip-5mm

\noindent
which is an asymptotically leading mode.
We remind that the solution (3) is not a general solution ,but only an
attractor
After the substitution of (15) into (12) one finds a Langevin equation
$$
 {d\varphi/ d\alpha} = -M^2{V'/ V} + \eta(\alpha)
\eqno(16)
 $$
where $ \eta(\alpha) $ is a white noise with the correlation function
$$
\VEV {\eta(\alpha)\eta({\alpha'})} = 2{D_0(\varphi)}{V(\varphi)\over
M^2}{1-{M^2
\eqno(17)
$$
where
$$  n\equiv n(\varphi) = 2({V\over MV'})^2 - 1,~~~~
 2D_0(\varphi) = {1\over 12{\pi}^2}{1\over \epsilon^{2\over n}}{(n + 2)^2\over
(
 \eqno(18)
  $$
 From the Langevin eq.(16) we obtain a Fokker-Planck equation
$$
 {\partial P\over \partial \alpha} = M^2 {\partial \over \partial
\varphi}({V'\o
 \eqno(19)
  $$
where
$$
 {A(\varphi)}^2 =D_0(\varphi){1 - {M^2\over 2}({V'\over V})^2\over {1 -
{M^2\ove
 \eqno(20)
  $$

As usual the Fokker-Planck equation (19) has two stationary solutions,one of
whi
$$
 P_0(\varphi) = {const\over A(\varphi)V^{1\over 2}}\exp[M^4\int {1\over
{A(\varp
 $$
corresponds to the "quantum creation" of the Universe since it has the
Hawking-M
$ P(\varphi)\approx {j_0V/ M^2V'}       $
and describes the "classical creation" of the Universe,i.e.,the evolution from
t

\centerline{Acknowledgements}

I am very grateful to Prof. Misao Sasaki and Dr. Ewan Stewart for very useful
discussions.
This work was supported in part by a JSPS Fellowship and by Monbusho Grant-
in-Aid for Encouragement of Young Scientists, No.92010.

\centerline{REFERENCES}

\noindent
Halliwell J. (1987)Scalar fields  in cosmology  with an exponential
 potential.
\nextline
{\it Phys.Lett.} B{\bf 185},341-344.

\noindent
Lucchin F.\& Matarrese S.(1985) Power-law inflation.
\nextline
{\it Phys.Rev.}D{\bf 32},1316-1322.

\noindent
Makino N.\& Sasaki M.(1991)The density perturbation in the chaotic
inflation  with non-minimal coupling.{\it Progr.Theor.Phys.}
{\bf 86},103-118.

\noindent
Ortolan A.,Lucchin F.\& Matarrese S.(1988)  Non-Gaussian
perturbations
\nextline
from inflationary dynamics.  { \it Phys.Rev.}D{\bf 38},465-471.

\noindent
Salopek D.S.\& Bond J.R.(1991)Stochastic inflation
and nonlinear gravity.\nextline
{\it Phys.Rev.} D {\bf 43}, 1005-1031.

\noindent
Sasaki M.,Nambu Y.\& Nakao K.(1988)Classical behavior of a scalar
field
\noindent
in the inflationary universe.{\it Nucl.Phys.}B{\bf308},868-884.

\noindent
Starobinsky,A.A.(1986) Stochastic de Sitter
  (inflationary)  stage in the
  early
\nextline
Universe.    {\it Lecture Notes in Physics}.{\bf 246}
,107-126  (\rm Proceedings
 of the seminar on
\nextline
 \it{Field Theory, Quantum Gravity
, and Strings},\rm Meudon
 and Paris VI,
\nextline
France 1984/85,
edited by
 H.J.de Vega and N.Sanchez)

\noindent
Yokoyama J.\& Maeda K.(1988) On the dynamics of the power law
 inflation due to exponential potential.{\it Phys.Lett.} B{\bf 207}
,31-35.

\end